\RequirePackage{fixltx2e}
\documentclass[aps,prb,reprint,floatfix,longbibliography]{revtex4-1}

\usepackage[bbgreekl]{mathbbol}
\usepackage{amsfonts}

\DeclareSymbolFontAlphabet{\mathbb}{AMSb}
\DeclareSymbolFontAlphabet{\mathbbl}{bbold}
\usepackage{mathtools} %
\usepackage{wasysym} %
\usepackage{graphicx} %
\usepackage[dvipsnames,table]{xcolor} %
\usepackage{tabularx} %
\usepackage[acronym]{glossaries} %
\usepackage{braket} %
\usepackage{fancyhdr} %
\usepackage{microtype} %
\usepackage{natmove}
\usepackage[byname]{smartref} %
\usepackage[breaklinks, %
colorlinks, linkcolor=Blue, citecolor=Blue, urlcolor=Blue]{hyperref} %
\usepackage[version=3]{mhchem}
\AtBeginDocument{\usepackage{booktabs}} 

\renewcommand{\Re}{\operatorname{Re}}

\newcommand{\eq}[1]{Eq.~(\ref{#1})}
\def\be{\begin{equation}} %
\def\ee{\end{equation}} %
\def\bea{\begin{eqnarray}} %
\def\eea{\end{eqnarray}} %

\newacronym[longplural={degrees of freedom}, %
firstplural={degrees of freedom (DOF)}, plural={DOF}]{DOF}{DOF}{degree
  of freedom} %
\newacronym[longplural={equations of motion}, %
firstplural={equations of motion (EOM)}, %
plural={EOM}]{EOM}{EOM}{equation of motion} %
\newacronym{PES}{PES}{potential energy surface} %
\newacronym{EH}{EH}{Ehrenfest}
\newacronym{SH}{SH}{fewest switches surface hopping}
\newacronym{MQC}{MQC}{mixed quantum-classical}
\newacronym{NAC}{NAC}{nonadiabatic coupling} %
\newacronym{SO}{SO}{split operator} %
\newacronym{LZ}{LZ}{Landau-Zener} %

\renewcommand{\v}[1]{\ensuremath{\mathbf{#1}}} 
 
\newcommand{\abs}[1]{\left| #1 \right|} 
\newcommand{\dpar}[2]{\frac{\partial #1}{\partial #2}}

\def\MyTitle{On the Breakdown of the Ehrenfest Method for Molecular Dynamics on Surfaces} %
\def\MyAuthora{Ignacio Loaiza} %
\def\MyAuthorb{Artur F. Izmaylov} %

\begin{document}

\title{\MyTitle}

\author{\MyAuthora{}} %
\affiliation{Chemical Physics Theory Group, Department of Chemistry,
  University of Toronto, Toronto, Ontario M5S\,3H6, Canada}
\affiliation{Department of Physical and Environmental Sciences,
  University of Toronto Scarborough, Toronto, Ontario, M1C\,1A4,
  Canada} %

\author{\MyAuthorb{}} %
\affiliation{Chemical Physics Theory Group, Department of Chemistry,
  University of Toronto, Toronto, Ontario M5S\,3H6, Canada}
\affiliation{Department of Physical and Environmental Sciences,
  University of Toronto Scarborough, Toronto, Ontario, M1C\,1A4,
  Canada} %

\date{\today}

\begin{abstract}
Due to a continuum of electronic states present in periodic systems, the description of molecular dynamics on
 surfaces poses a serious computational challenge. One of the most used families of approaches in these settings 
 are friction theories, which are based on the \gls{EH} approach. Yet, a mean-field treatment of electronic degrees 
 of freedom in the \gls{EH} method makes this approach inaccurate in some cases. Our aim is to clarify when \gls{EH} breaks down for molecular dynamics on surfaces. Answering this question provides limits of applicability 
 for more approximate friction theories derived from \gls{EH}. We assess the \gls{EH} method on one-dimensional, numerically exactly solvable models with a large but finite number of electronic states. Using the Landau-Zener formula 
 and the Massey parameter, an expression that determines when \gls{EH} breaks down is deduced.
\end{abstract}

\glsresetall

\maketitle

\section{Introduction}
Molecular dynamics in periodic systems in general and 
on metallic surfaces in particular is complicated by the presence of a continuum of electronic states. 
This continuum can lead to break-down of the Born-Oppenheimer approximation, 
which is based on the assumption that the energy separation between electronic states is much larger than  
the nuclear kinetic energy.  
This means that a proper treatment of such systems must take into account nonadiabatic effects. 
The most computationally feasible candidates for simulations in extended systems are 
\gls{MQC} methods~\cite{ilya, metal_fssh, shenvi,Gherib:2015ix}, such as the \gls{EH} approach~\cite{doltsinis}, 
and Tully's \gls{SH} algorithm~\cite{fssh,doltsinis,Wang:2016bza}. 
Yet, due to the continuum of electronic states, \gls{EH} and \gls{SH} are not straightforwardly applicable to periodic systems. First, the continuum of states must be discretized~\cite{discretization}, 
and even then thousands of electronic states can be involved in dynamics,\cite{iesh} rendering such calculations computationally challenging. 

To avoid explicit treatment of a large number of electronic states, frictional theories were introduced. In these theories, 
starting from the \gls{EH} formalism, the electronic continuum is approximately integrated,\cite{fric}
and nuclear trajectories evolve on the adiabatic ground state 
with an additional frictional force arising from nonadiabatic transitions.   
The frictional approaches have been extensively 
studied over the years \cite{fric,tens_fric,mode_fric,dou_fric,many_body_fric,q_fric}. 
The problem with frictional approximations appears when multiple electronic surfaces with very different 
nuclear dependence become energetically available. This often takes place for different 
adsorbate states, for example, dynamics of $\mathrm{Cl}$/$\mathrm{Cl}^-$ adsorbates on the Cu (110)
surface in scanning tunneling 
microscope induced reactions.\cite{polanyi, polanyi2} Here,
the underlying \gls{EH} may not be as good as in the case when only one adsorbate state is involved. 

Some inaccuracies introduced by frictional approximations were corrected by switching to 
a collective variable approach~\cite{ilya}. This approach does not treat electronic DOF fully implicitly  
but rather reduces their explicit consideration to only few (two or three) collective electronic coordinates. 
However, in the current formulation this approach still cannot surpass the EH method which it is based on.
Applying the collective variable idea to the \gls{SH} framework would be ideal in cases when several adsorbate 
states are involved in surface dynamics, since this framework allows for proper account 
of dynamics on individual electronic states~\cite{Loaiza2:inprep}. 

Still, \gls{EH} and \gls{EH}-based methods are frequently used in modeling surface dynamics, and 
it is instructive to develop a numerical criterion indicating when going beyond the EH framework is necessary. This is the main objective of the present work. We begin by reviewing the EH and SH methods with an illustration of EH failure  
on Tully's extended coupling model in Sec.~\ref{subsec:MQC}. Using the \gls{LZ} formula, we deduce a parameter that can indicate when \gls{EH} dynamics breaks down in Sec.~\ref{subsec:EHBI}. Section \ref{sec:sim} illustrates
 performance of the new indicator on one-dimensional surface dynamics inspired models.

\section{Theory} \label{sec:theory}
\subsection{Mixed quantum-classical methods} \label{subsec:MQC}

To introduce notation and to provide self-sufficient description of methodology we start with a 
brief reminder of the \gls{EH} and \gls{SH} methods.\cite{fssh,doltsinis,tullymqc}
In both of these methods, nuclear coordinates $\v R$ are treated classically and are substituted by corresponding 
time-dependent functions $\v R=\v R(t)$. Using this substitution one can formulate the total non-stationary electronic 
wavefunction as 
\begin{equation} \label{eq:ansatz}
\ket{\Psi(\v R(t),t)}=\sum_{j}c_j(t)\ket{\phi_j[\v R(t)]},
\end{equation}
where $\ket{\phi_j[\v R(t)]}$ are eigenfunctions of the electronic Hamiltonian $\hat H_e[\v R(t)]$,
$\hat H_e[\v R(t)]\ket{\phi_j[\v R(t)]}=E_j[\v R(t)]\ket{\phi_j[\v R(t)]}$, and $E_j[\v R(t)]$ are corresponding 
adiabatic PESs. To formulate equations of motion for time-dependent coefficients $c_j(t)$ 
the time-dependent Schr\"odinger equation 
\begin{equation} \label{eq:TDSE}
i\dpar{}{t}\ket{\Psi(\v R(t),t)}=\hat H_e[\v R(t)]\ket{\Psi(\v R(t),t)}
\end{equation}
is projected onto the adiabatic wavefunctions $\ket{\phi_j[\v R(t)]}$ 
\begin{equation} \label{eq:elec_dyn}
\dot{\v c}=-(i\v{H_e} +\dot{\v R} \cdot \v\Gamma)\v c,
\end{equation}
where $\v\Gamma$ is a matrix with elements $\v\Gamma_{jk}=\braket{\phi_j[\v R] | \nabla_{\v R} \phi_k[\v R]}$ corresponding to the \gls{NAC} vectors, and $\v c$ is a vector with entries $c_j(t)$. 
In the adiabatic representation, $\v{H_e}$ is a diagonal matrix with elements 
$\v{H_e}_{,kj}=\braket{\phi_k(\v R) | \hat H_e (\v R) | \phi_j(\v R)}=E_k (\v R)\delta_{kj}$.

In \gls{EH} and \gls{SH} methods the nuclear dynamics, $\v R(t)$, is governed by Newton equations of motion. 
For \gls{EH}, the force is based on the gradient of the averaged electronic energy
\begin{equation}
\v F=-\nabla_{\v R}\braket{\Psi | \hat H_e | \Psi}.
\end{equation}
Using the adiabatic expansion for $\ket{\Psi}$ in \eq{eq:ansatz} and 
\bea
\nabla_{\v R}\braket{\phi_j | \hat H_e | \phi_k}=(E_k-E_j)\v\Gamma_{jk}+\braket{\phi_j | \nabla_{\v R}\hat H_e | \phi_k}
\eea
one obtains the nuclear equation of motion in the Ehrenfest method as
\begin{equation}
M\ddot{\v R}=-\sum_k \vert c_k\vert^2\nabla_{\v R} E_k+\sum_{k\neq j}c_k^*c_j(E_k-E_j)\v\Gamma_{kj}.
\end{equation}
This equation can also be derived from the conservation of the total energy $E=\frac{1}{2}M \abs{\dot{\v R}}^2+\braket{\Psi | \hat H_e | \Psi}$.

In \gls{SH}, nuclear dynamics is governed by forces from a single \gls{PES} at each moment of time
\begin{equation}
M\ddot{\v R}=-\nabla_{\v R} E_j(\v R).
\end{equation}
However, to allow for nonadiabatic dynamics, a probability $P_{j\rightarrow k}$ of
changing a \gls{PES} corresponding to a state $\ket{\phi_j}$ to that of a state $\ket{\phi_k}$
is introduced for every trajectory, a so-called hopping probability,
\begin{equation} \label{eq:hop_prob}
P_{j\rightarrow k}=\frac{2\Re (c_j^*c_k\dot{\v R}\cdot \v\Gamma_{jk})\Delta t}{\vert c_j \vert ^2},
\end{equation}
where $\Delta t$ is a time-step. 
In both methods an ensemble of trajectories is used to model the quantum nuclear distribution. 

\begin{figure}[]
  \centering
    \includegraphics[width=7cm]{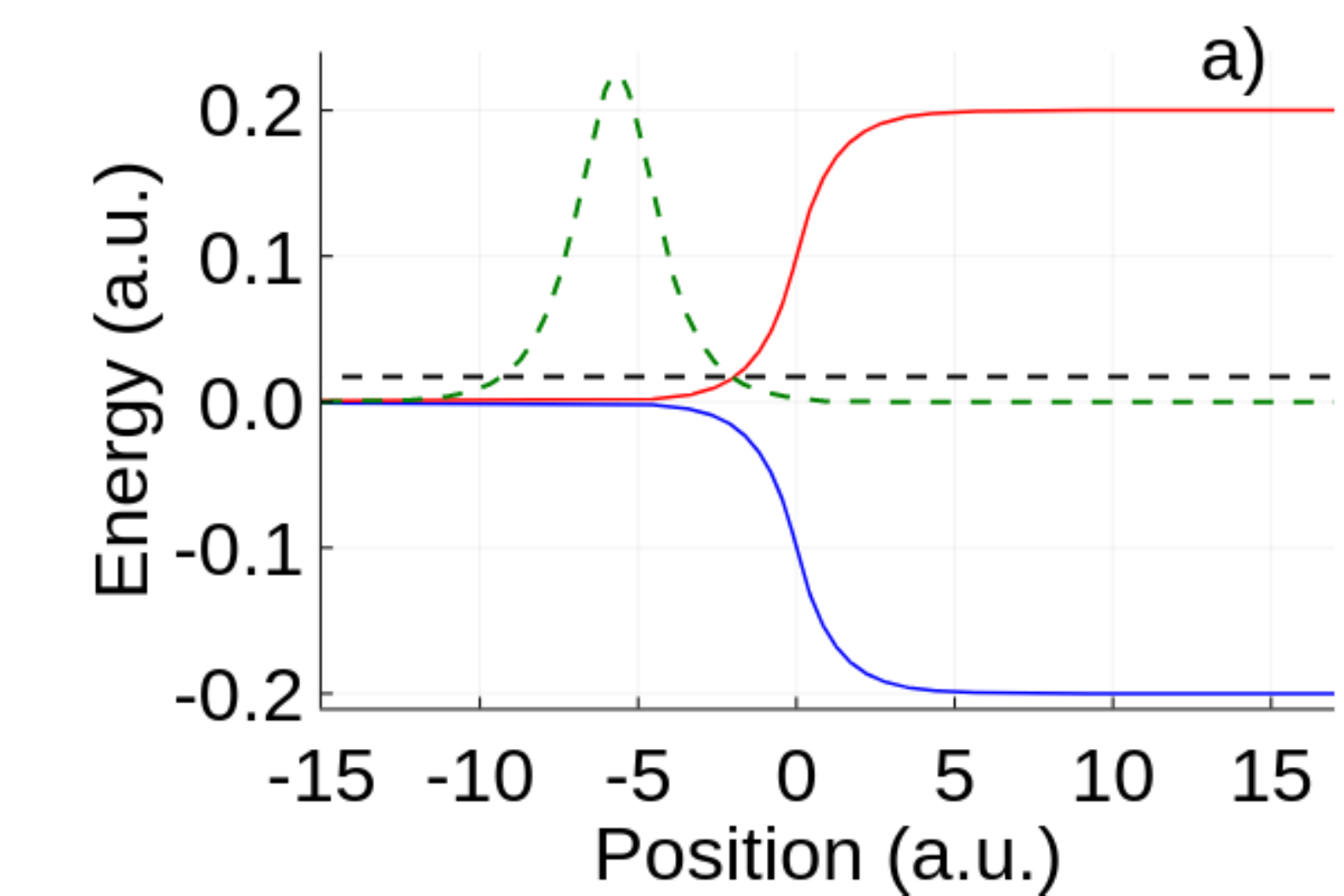}
    \includegraphics[width=7cm]{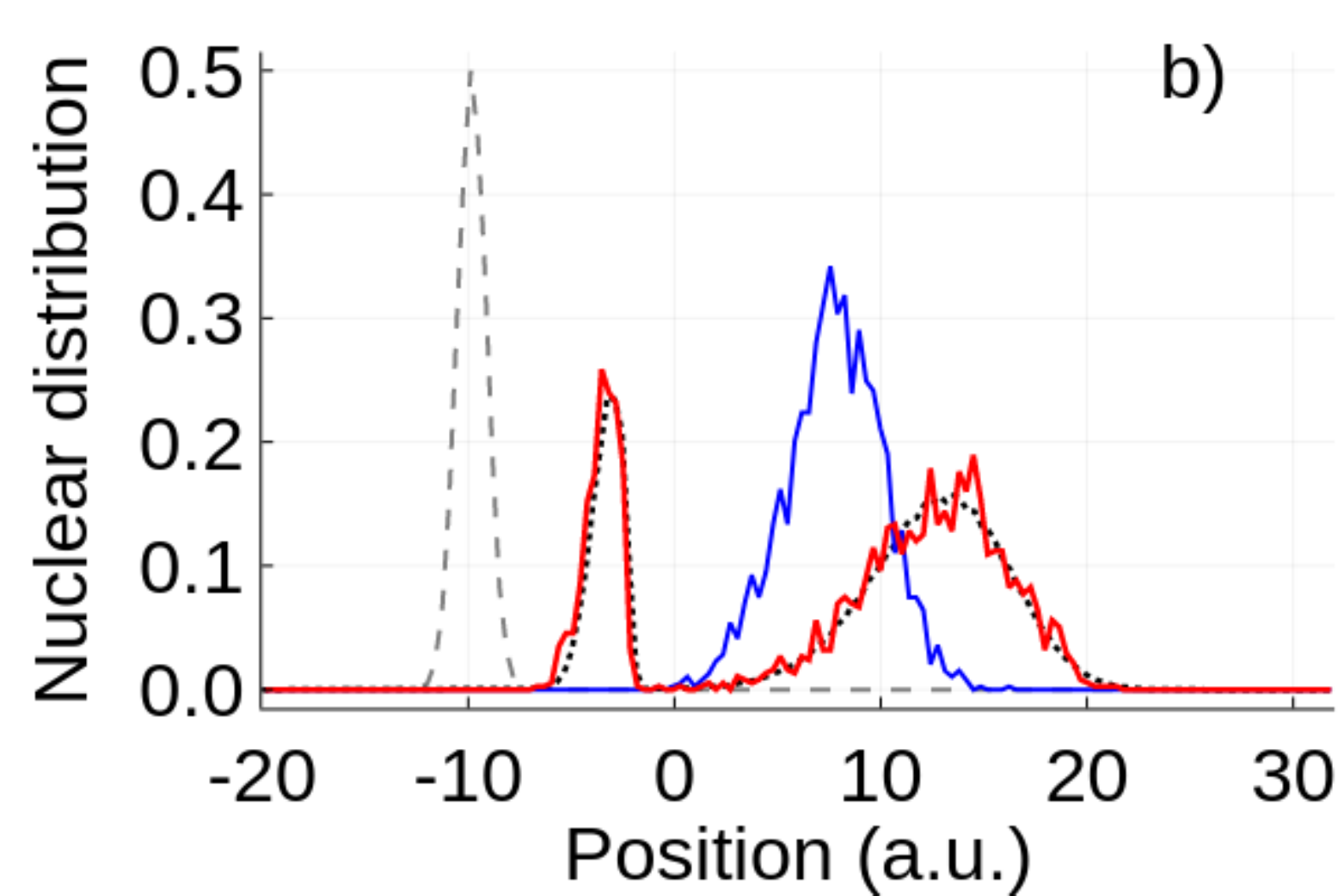}
  \caption{a) \glspl{PES} (red and blue solid) for Tully's extended coupling model \cite{fssh}, the initial energy (horizontal dashed), and \glspl{NAC} (green dashed). b) Initial (dashed gray, its magnitude is divided by $2$) and final nuclear position distributions (blue for \gls{EH}, red for \gls{SH}, and black dots for exact quantum). The initial Wigner distribution is centered at $R=-10$ and has average momentum $p=8.5$.}
  \label{fig:extended}
\end{figure}

Thus, \gls{EH} and \gls{SH} have the same electronic dynamic equation Eq.\eqref{eq:elec_dyn}, but different 
nuclear dynamics. These differences in treatments of nuclear DOF can result in significant differences in nuclear dynamics in cases where the following two conditions are satisfied: 
1) the coupling region is followed by the region where participating PESs have different slopes, 
2) probabilities to find the system on either of participating PESs are similar. 
A good, illustrative example of these differences 
is dynamics in the Tully extended coupling model \cite{fssh} (Fig.~\ref{fig:extended}a). 
The difference in PES slopes after the coupling region and the chosen initial energy
causes the nuclear wave-packet to split, 
the part on the excited \gls{PES} is reflected while that on the ground state \gls{PES} is transmitted (Fig.~\ref{fig:extended}b). 
\gls{SH} gives a nuclear distribution almost indistinguishable from the exact quantum one, 
while \gls{EH} fails to capture the branching of the nuclear distribution. 
For the comparison with the exact quantum dynamics, we used the \gls{SO} method.\cite{tannor}.

\subsection{Ehrenfest break-down indicator} \label{subsec:EHBI}

To provide more quantitative measure for possibility of the Ehrenfest method failure in situations where 
involved PESs have different slopes, we will use the \gls{LZ} formula~\cite{zener} 
to estimate the probabilities of finding the system on different PESs. The \gls{LZ} equation does not require 
coefficients for the electronic states entering \eq{eq:hop_prob}, and therefore, is more convenient for estimates. 
Of course, use of the \gls{LZ} expression introduces certain constraints,
however, such constraints are generally consistent with considered processes of interest: molecular dynamics of an 
adsorbate that has at least two different molecular electronic states, whose PESs approach each other 
in localized nuclear configuration regions. Thus, all models considered in this work have  
nonadiabatic couplings localized near crossings of diabatic potentials. 

\begin{figure}[]
  \centering
    \includegraphics[width=7cm]{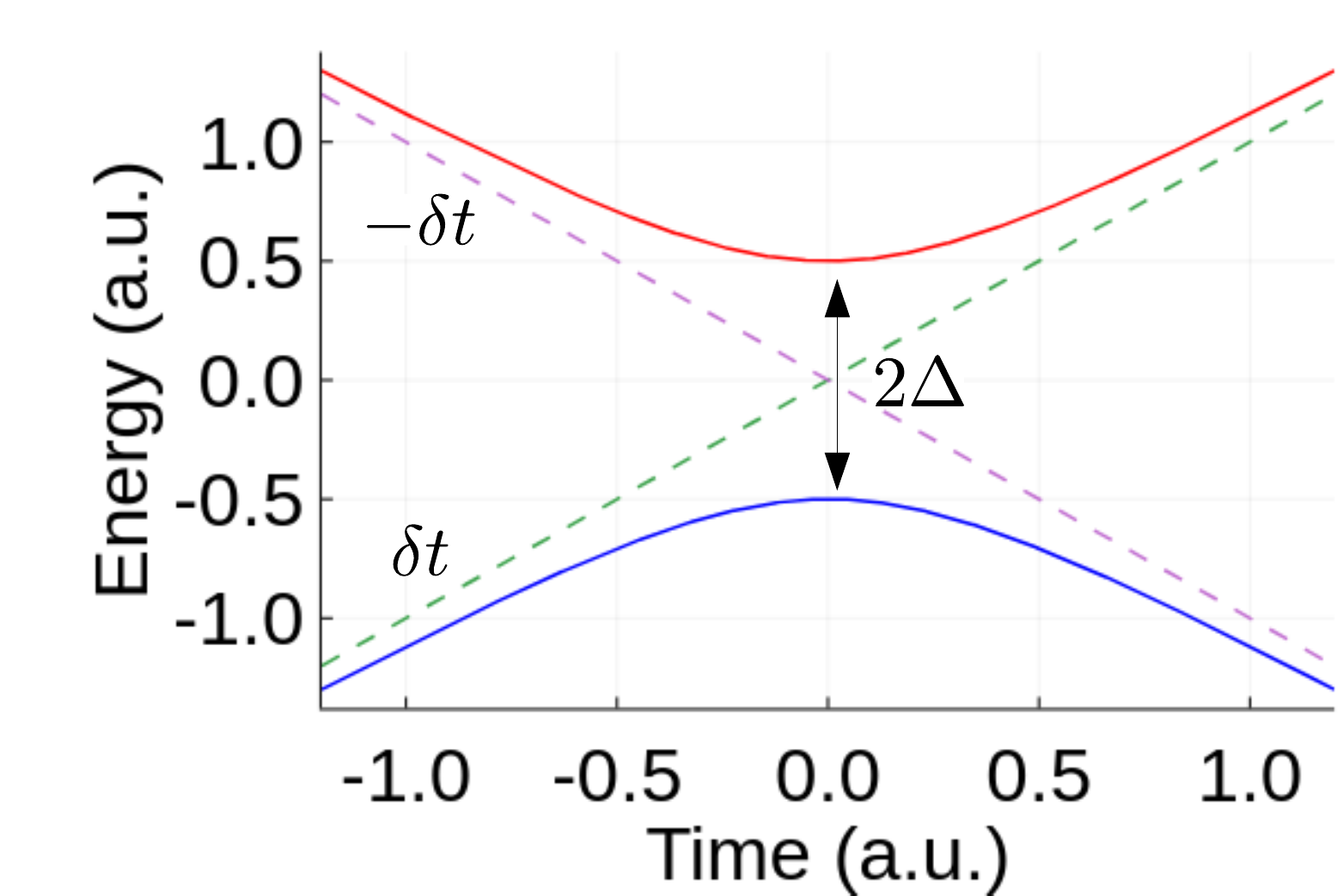}   
  \caption{Landau-Zener crossing model with $\delta=1$ and $\Delta=0.5$. Adiabatic (diabatic) \glspl{PES} are solid (dashed) lines.}
  \label{fig:LZ}
\end{figure}

The \gls{LZ} potential assumes that the transition region is so small that the energy difference between diabats may be seen as a linear function in time ($2\delta t$), while the off-diagonal coupling $\Delta$ is a constant, and is defined, along with all its auxiliary variables, in Appendix A. 
The \gls{LZ} formula for calculating the probability of changing from diabatic surface $a$ to $b$ 
after passing through a diabatic potentials (Fig.~\ref{fig:LZ}) crossing point is

\begin{equation} \label{eq:LZ}
P_{b\leftarrow a}=1-\exp\Big ({\frac{-2\pi \Delta^2}{\dot R \abs{F_b-F_a}}} \Big ),
\end{equation}
where $\dot R$ is the nuclear velocity evaluated at the crossing point, and 
$F_b$ and $F_a$ are the diabatic forces at that point.
In the adiabatic picture, $P_{b\leftarrow a}$ corresponds to the probability of staying on the adiabatic \gls{PES} 
that was very similar to the diabatic potential of the initial diabatic state ($a$) and ($b$).

The argument of the exponential function is called the Massey parameter\cite{massey}
\begin{equation} \label{eq:massey_d}
\xi_{\rm dia}=\frac{2\pi\Delta^2}{\dot R\abs{F_b-F_a}}.
\end{equation}
It can be used as an indicator for adiabatic ($\xi \gg 1$) or diabatic ($\xi \ll 1$) behavior. It was originally derived 
 in the diabatic basis~\cite{zener}, but it can also be obtained in the adiabatic representation using a perturbative approach~\cite{adiabatic_lz}. Also, using the LZ model, it is possible to transform the Massey parameter 
 $\xi_{\rm dia}$ to the adiabatic representation (see Appendix A for details) 
\begin{equation} \label{eq:massey_a}
\xi_{\rm adi}=\frac{\pi(E_2-E_1)}{4\abs{\braket{\phi_1|\partial_t\phi_2}}}=\frac{\pi(E_2-E_1)}{4\dot R \abs{\Gamma_{12}}},
\end{equation}
where $E_i$'s are adiabatic energies, $\braket{\phi_1|\partial_t\phi_2}$ is a time-derivative coupling, and $\Gamma_{12}$ is the \gls{NAC}; all 
quantities are taken at the diabatic crossing point or at the minimal gap point. 
The adiabatic formulation is beneficial for analyzing the data from first-principles calculations that do not 
have underlying diabatic models. Because of this, we will use $\xi\equiv\xi_{\rm adi}$.

From previous consideration we know that \gls{EH} will be
problematic when probabilities of finding the system on
competing pathways are similar, we judiciously consider 
the Massey parameter $\xi =1$ to be an indicator of this case. 

To calculate Massey's parameter from either \eq{eq:massey_d} or \eq{eq:massey_a}, the value of the nuclear velocity 
can be estimated from the energy conservation condition. We have two different values for $\dot R$: $\dot R_{\rm adi}$, which assumes a motion on the ground adiabatic state, and $\dot R_{\rm dia}$ for a motion in the first diabatic state. Both velocities would then represent an unperturbed motion in their respective bases.
Given an initial total energy $\epsilon$ for a trajectory, the velocities are evaluated as
\begin{equation}\label{eq:Rdot}
\dot R=\Big ( \frac{2}{M}(\epsilon-E)\Big )^{\frac{1}{2}},
\end{equation}
where for $\dot R = \dot R_{\rm dia} (\dot R_{\rm adi})$ we use $E=E_a(E_1)$ at the diabatic crossing point.
Since we are working with the adiabatic values, we will be using the velocity estimate using 
$E = E_1$ in \eq{eq:Rdot}.

\section{Results and Discussion} \label{sec:sim}

To illustrate correlation between our indicator and performance of mixed quantum-classical (MQC) methods we performed simulations with the EH, SH, and SO methods on two one-dimensional models.  
All MQC simulations were run using $2000$ trajectories with initial conditions sampled from a Wigner Gaussian distribution with the standard deviation of $1/\sqrt{2}$ for both nuclear positions and momenta. Dynamics were propagated using the fourth-order Runge-Kutta method for $\approx 121$ fs, with a time-step of $\Delta t=1.0 (0.1)$  a.u. for a sinusoidal (metallic surface) model. 
In all cases, dynamics were checked to yield converged results with respect to the time-step. The particle mass was set to be of $2000$ a.u.

All \gls{SH} dynamics were done in the adiabatic basis except in the low coupling case shown in Fig.~\ref{fig:sinus}c,d. For this case, the motion is highly diabatic, and as discussed in Ref.~\citenum{diavsadia}, \gls{SH} yields better results in the representation with fewer hops, which is the diabatic representation in this case.

All coding was done using the Julia language, and the code 
is available at \url{https://github.com/iloaiza/MQC}.

\subsection{Sinusoidal model}

A sinusoidal model has a diabatic potential of the form
\begin{equation}
V(R)=\left[ {\begin{array}{cc}
A\sin (kR) & \Delta \\
\Delta & -A\sin (kR) \\
\end{array} } \right].
\end{equation}
This potential will serve to model periodic systems, while offering a very simple parametric dependence 
that allows us to easily explore several coupling regimes by varying $\Delta$ with fixed $A=0.02$ and $k=0.5$. 

Based on particle's initial energy we distinguish the following dynamical regimes in this model:
\mbox{(1) Trapped regime}: the initial energy is lower than the adiabatic potential energy barrier height on the ground adiabatic state. 
This case is trivial since there is no appreciable non-adiabatic dynamics taking place, instead the particle is 
trapped on the ground surface site. 
(2) Adiabatic regime: there is enough energy to cross the adiabatic barrier but not  
for appreciable probability of nonadiabatic transition to the excited state. 
(3) Nonadiabatic regime: the particle has enough energy 
to be promoted to the excited state but not enough to cross the excited state barrier. 
(4) High energy regime: the initial energy is higher than the excited \gls{PES} barrier. 
For our investigation only the second, third and fourth regimes are of interest. 

\begin{figure*}[]
  \centering
    \includegraphics[width=7cm]{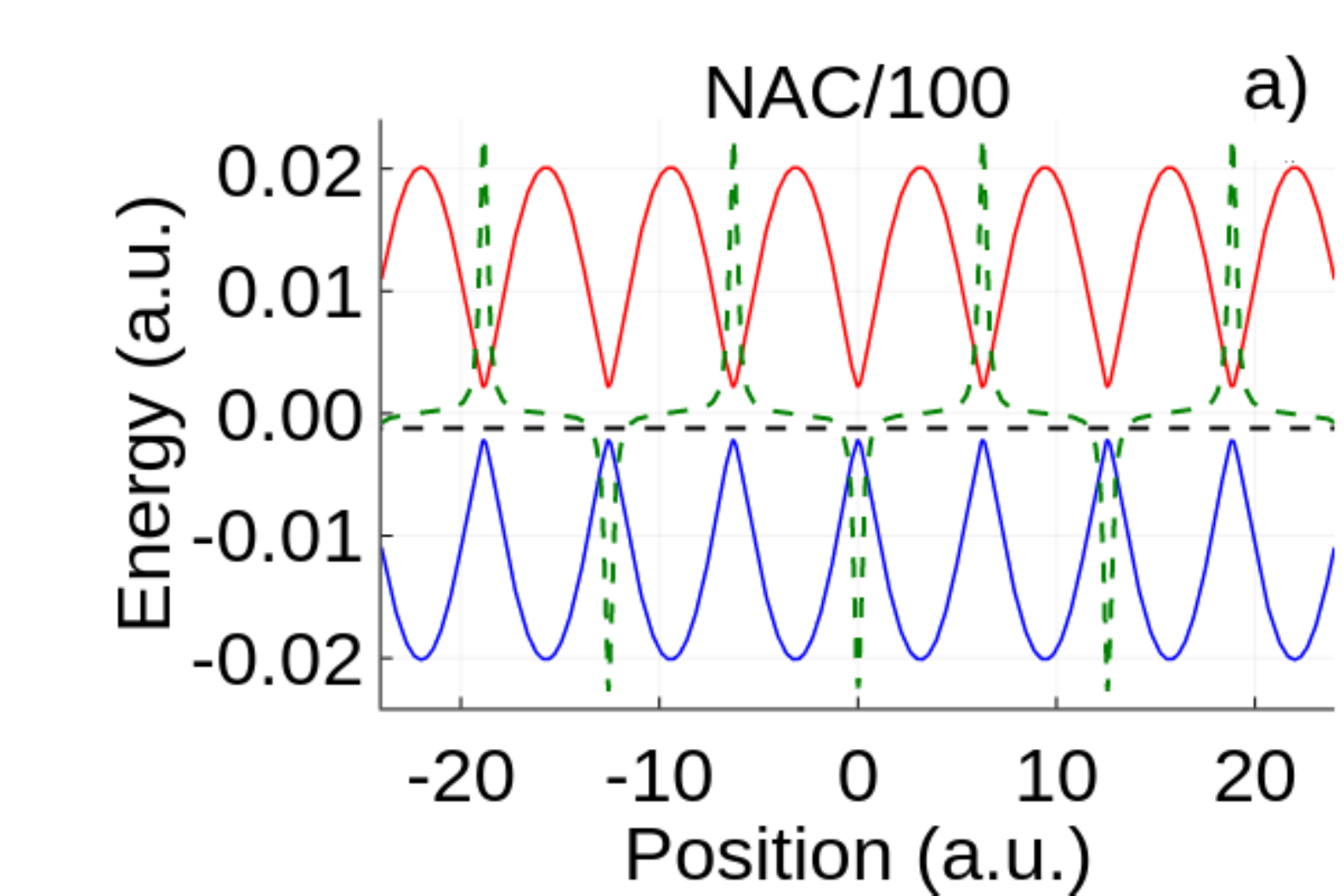}
    \includegraphics[width=7cm]{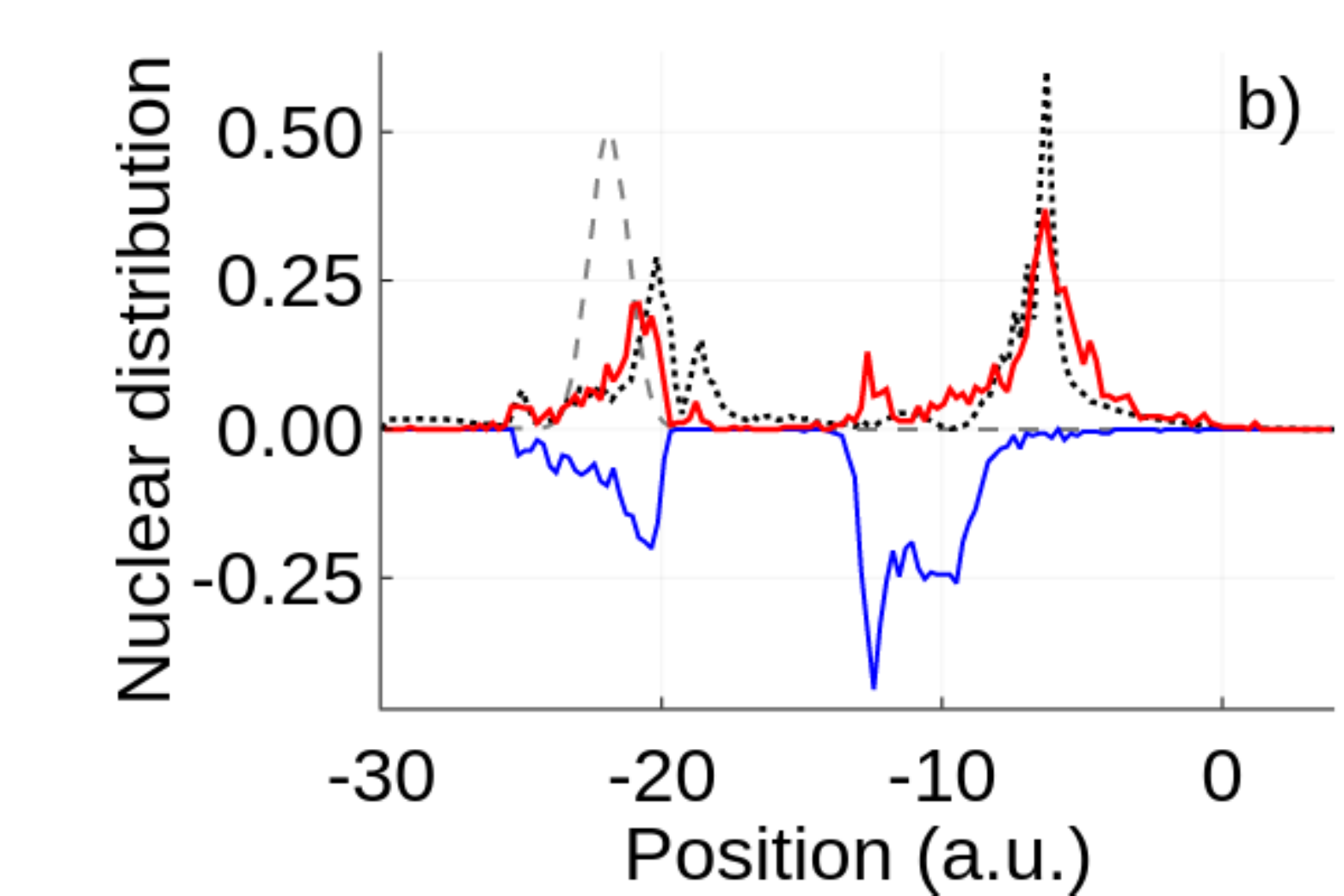}
  \caption{a) Sinusoidal model with coupling $\Delta=0.0022$ and $\xi=1.5$: \glspl{PES} (solid), initial energy (horizontal dashed), and \gls{NAC} (green dashed). b) Initial (dashed gray, its magnitude divided by $2$, the average momentum $p=8.7$) and final nuclear positions for nuclear trajectories (blue for \gls{EH}, red for \gls{SH}, and black dots for SO).   \gls{EH} results were mirrored for clarity.}
  \label{fig:sinus_0022}
\end{figure*}

\paragraph{Adiabatic regime:} Figure ~\ref{fig:sinus_0022} shows the sinusoidal potential and the nuclear distributions for all three methods. Even when the energy is not enough to get to the excited adiabatic state, by tuning the coupling so that $\xi\approx 1$ we create a population in the excited state, making \gls{EH} dynamics slower; whereas \gls{SH} dynamics will experience rejected hops, following a trajectory on the ground state that reproduces the \gls{SO} results accurately.

\begin{figure*}[]
  \centering
    \includegraphics[width=7cm]{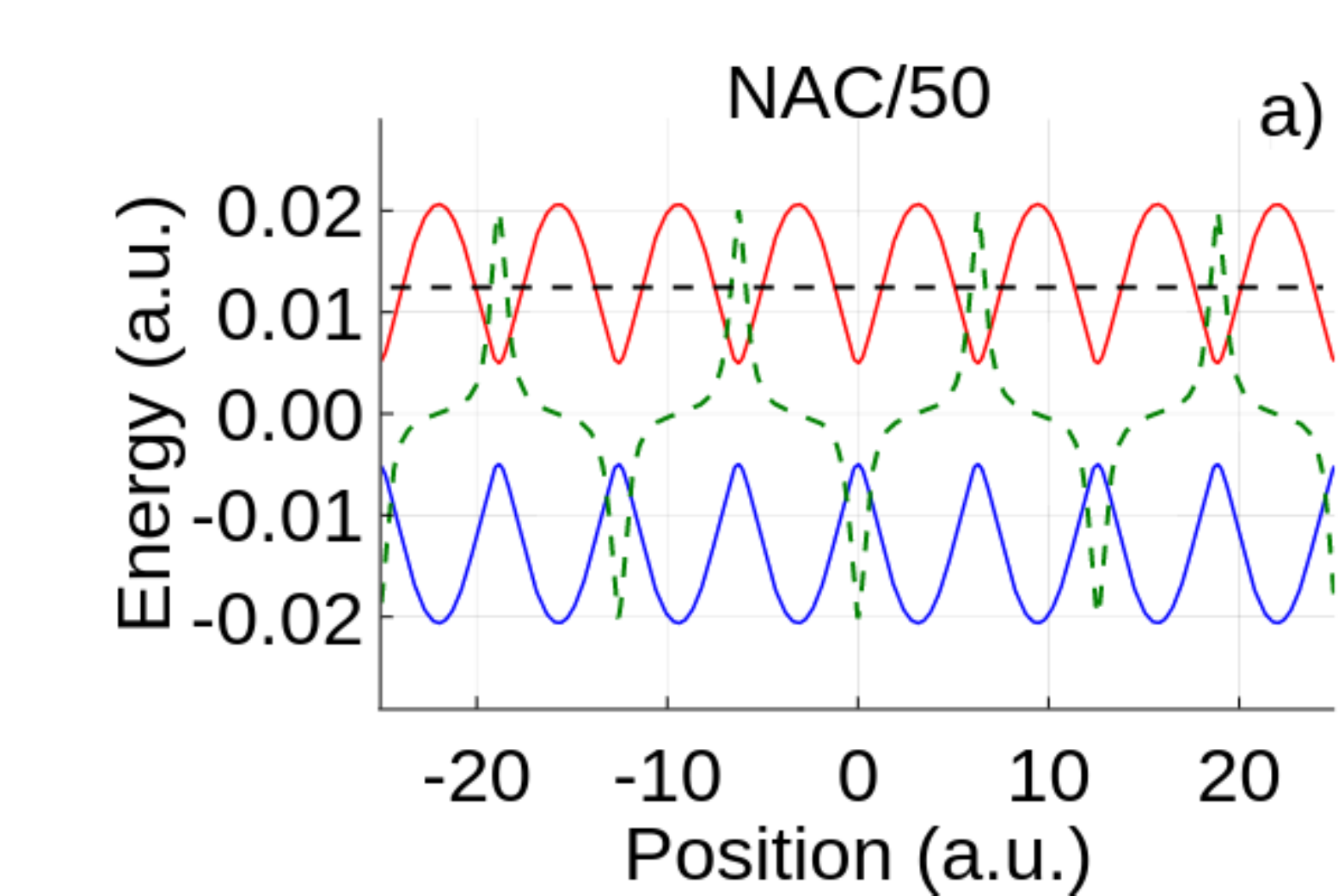}
    \includegraphics[width=7cm]{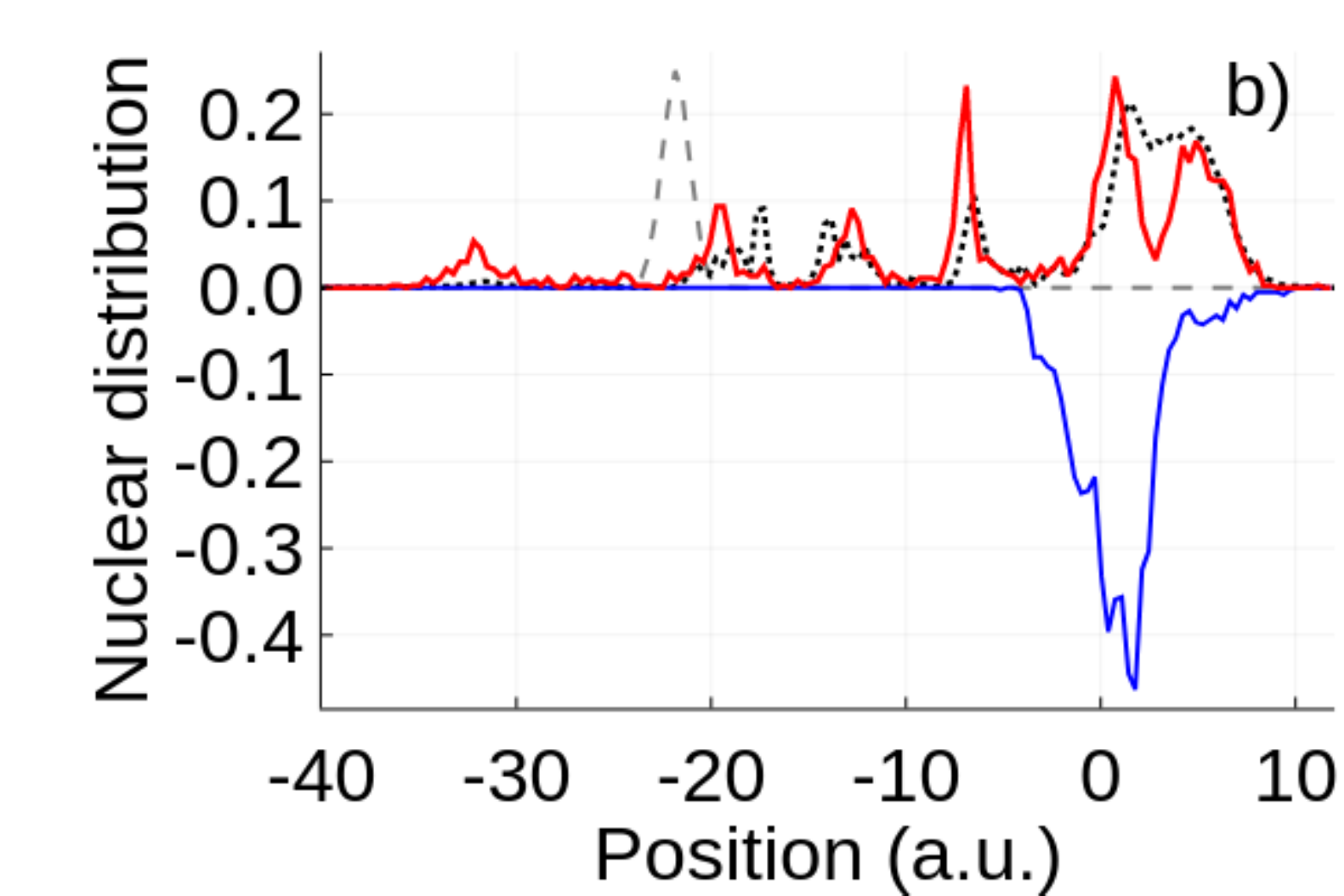}
    \includegraphics[width=7cm]{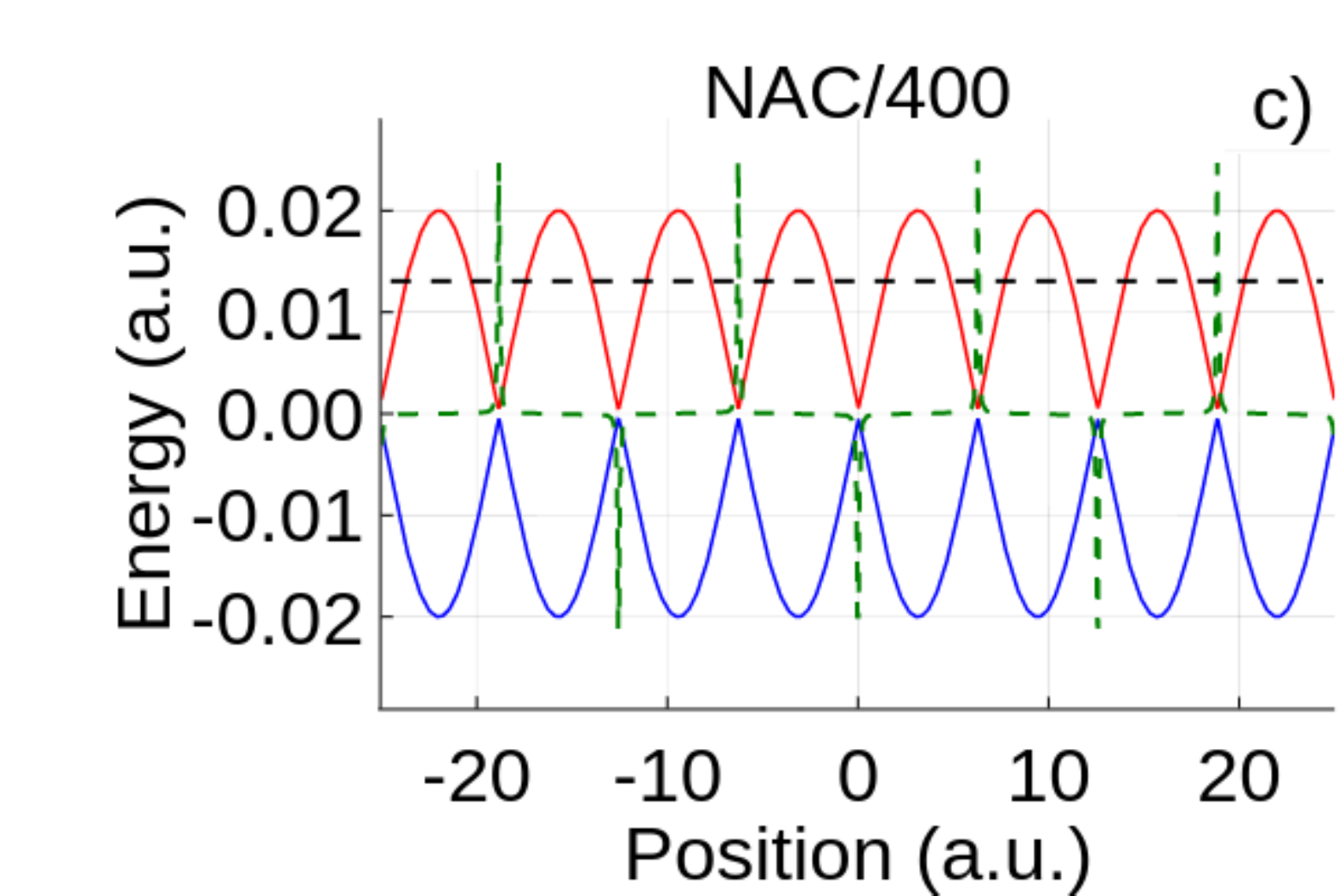}
    \includegraphics[width=7cm]{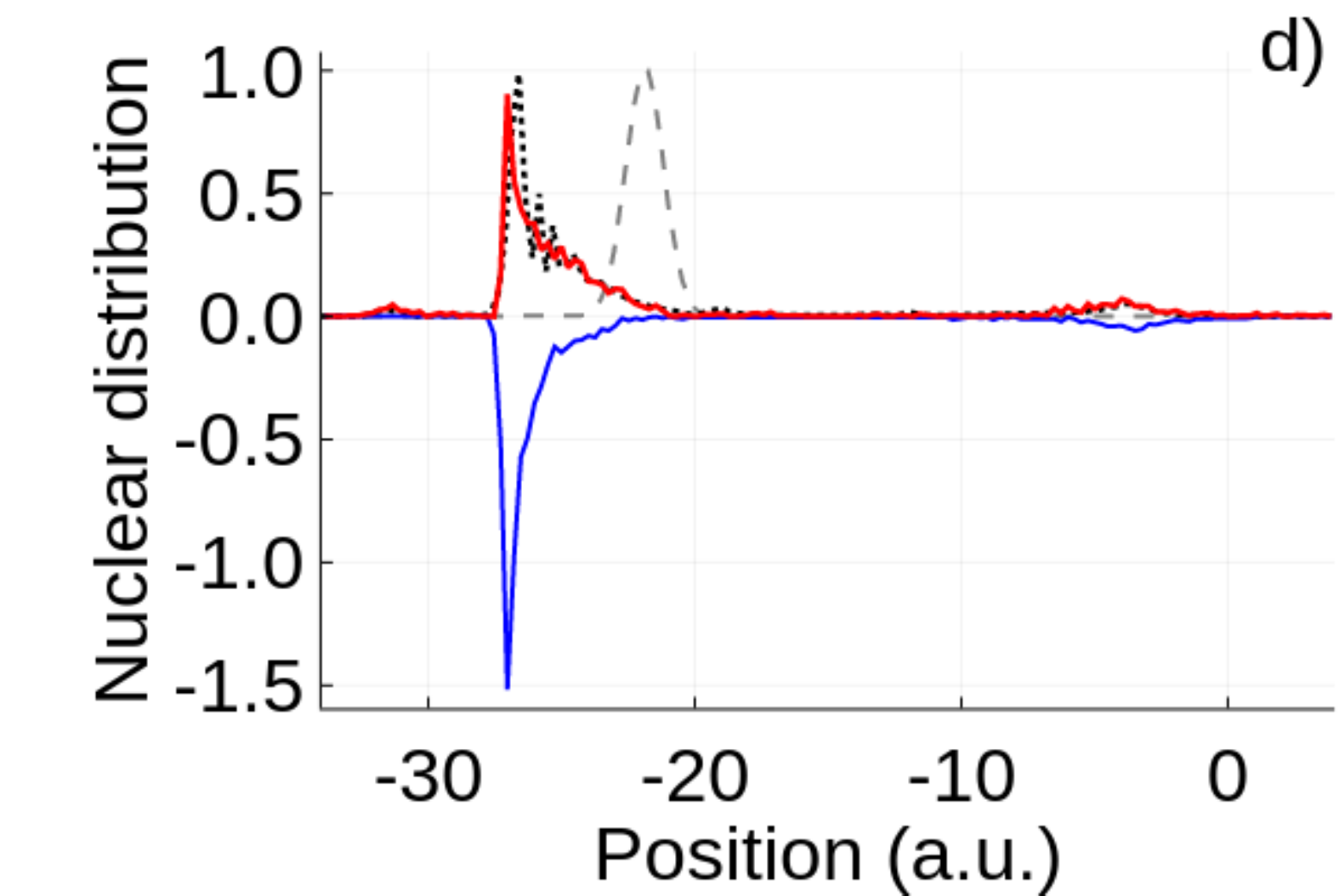}  
	\includegraphics[width=7cm]{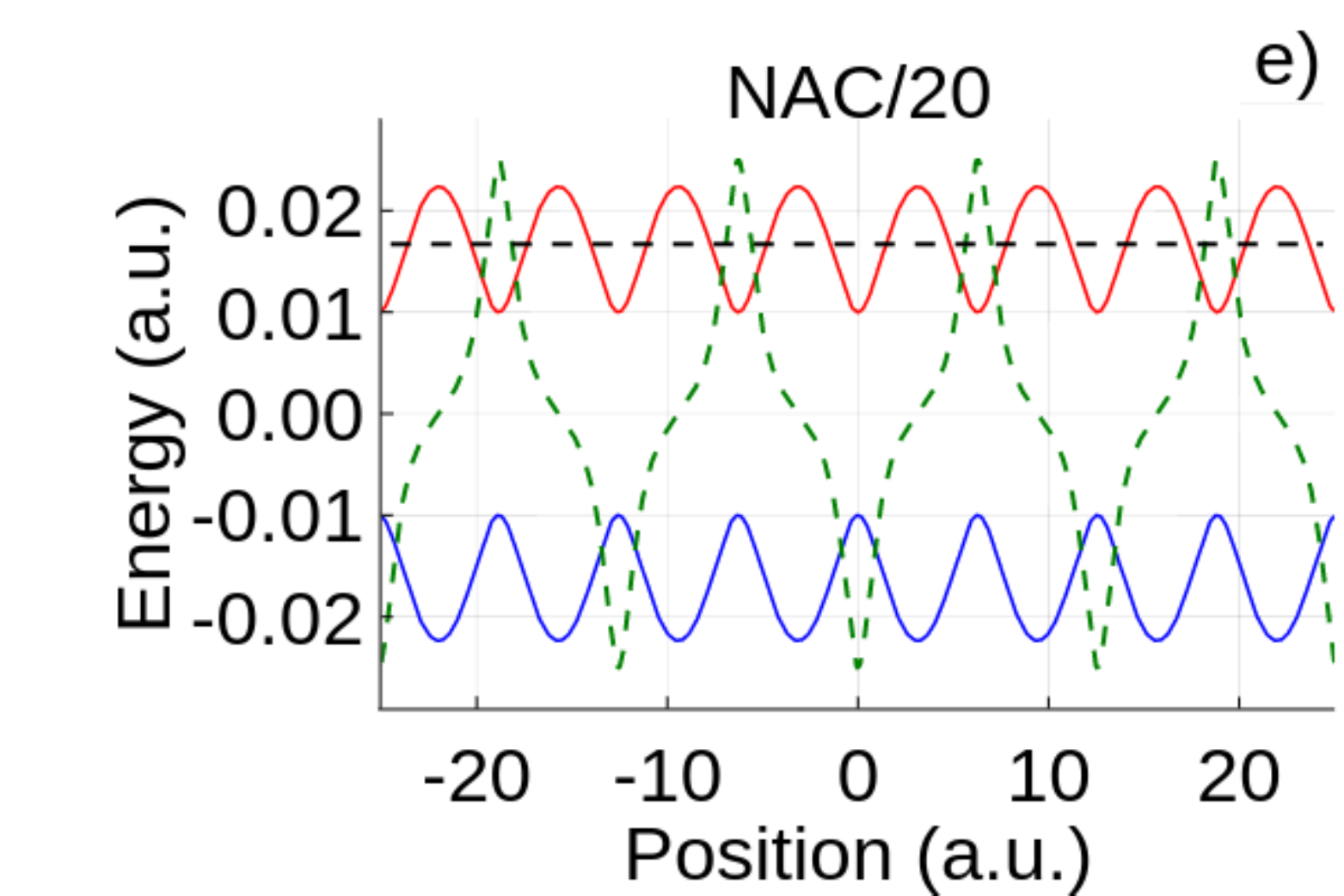}
    \includegraphics[width=7cm]{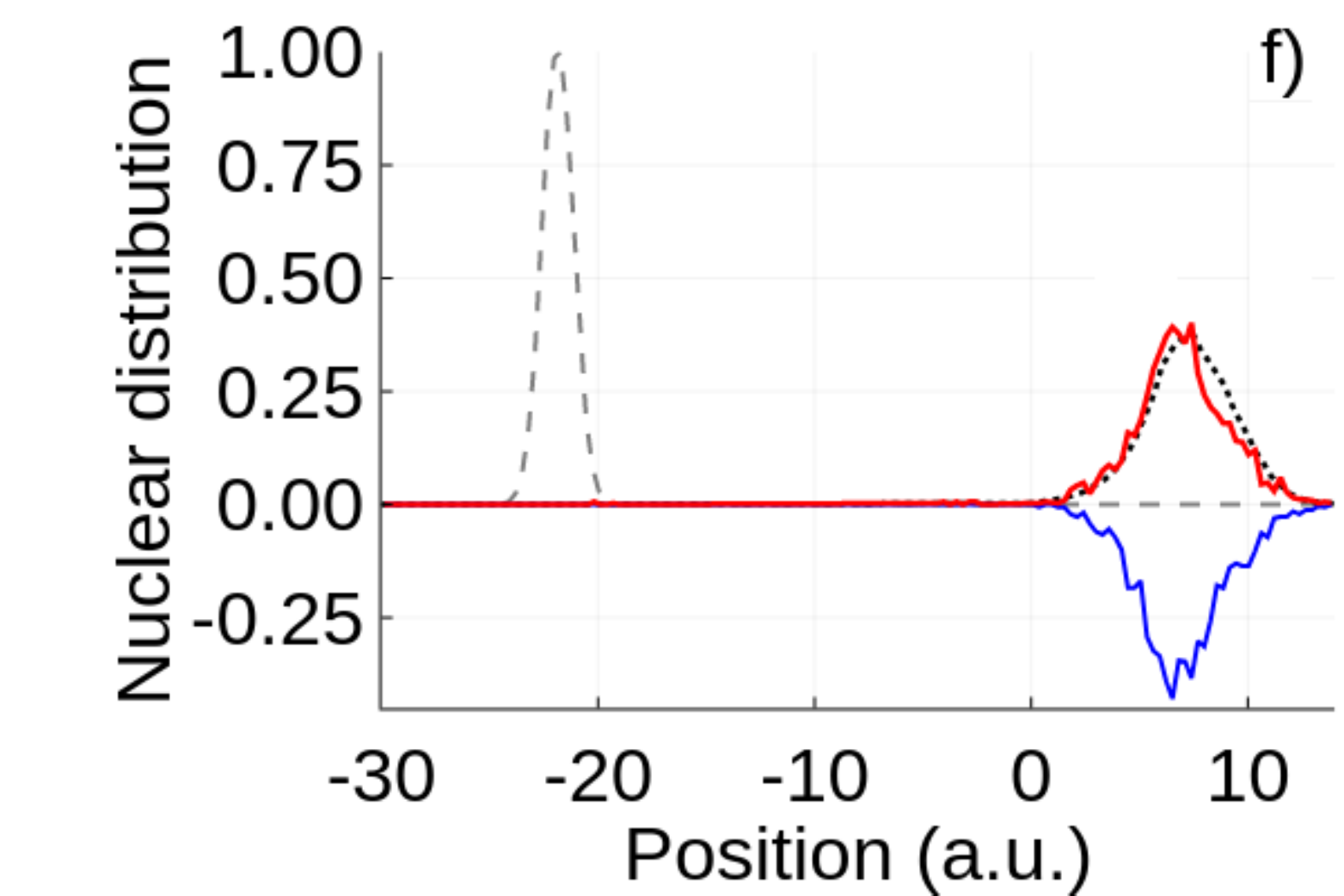}   
  \caption{a), c) and e): \glspl{PES} (solid), initial energy (horizontal dashed), and \glspl{NAC} (green dashed). b), d) and f): initial (dashed gray, its magnitude divided by $4$ for b) and final nuclear distributions (blue for \gls{EH}, red for \gls{SH}, and black dots for SO). Diabatic couplings, $\Delta$: $0.005$ (b), $0.0005$ (d), and $0.01$ (f). Initial Wigner distributions' average momenta: $p_0=11.5$ (b and d) and $p_0=12.5$ (f). \gls{EH} results were mirrored for clarity. The adiabatic Massey parameters, $\xi$: $1.9$ (b), $0.021$ (d), and $6.1$ (f).}
  \label{fig:sinus}
\end{figure*}

\paragraph{Nonadiabatic regime:}Figure \ref{fig:sinus} shows nuclear distributions for the different coupling regimes. 
All three methods yield very similar results if Massey's parameter is not near $1$. 
For large and small couplings, the nuclear wave-packet moves either highly adiabatically or diabatically 
and both \gls{EH} and \gls{SH} reproduce quantum simulations well (Figs.~\ref{fig:sinus}d,f).  

Branching of nuclear trajectories will be particularly noticeable for the intermediate coupling case that corresponds 
to $\xi=1.9$ (Fig.~\ref{fig:sinus}a). At every diabatic crossing, the coupling splits the ensemble into two parts, 
one is following the adiabatic route on the ground state and the other is transferred to the upper state and is reflected by 
its repulsive part. These competing pathways with different behaviors are 
not captured well with \gls{EH} dynamics (Fig.~\ref{fig:sinus}b). 

\begin{figure}[]
  \centering
    \includegraphics[width=7cm]{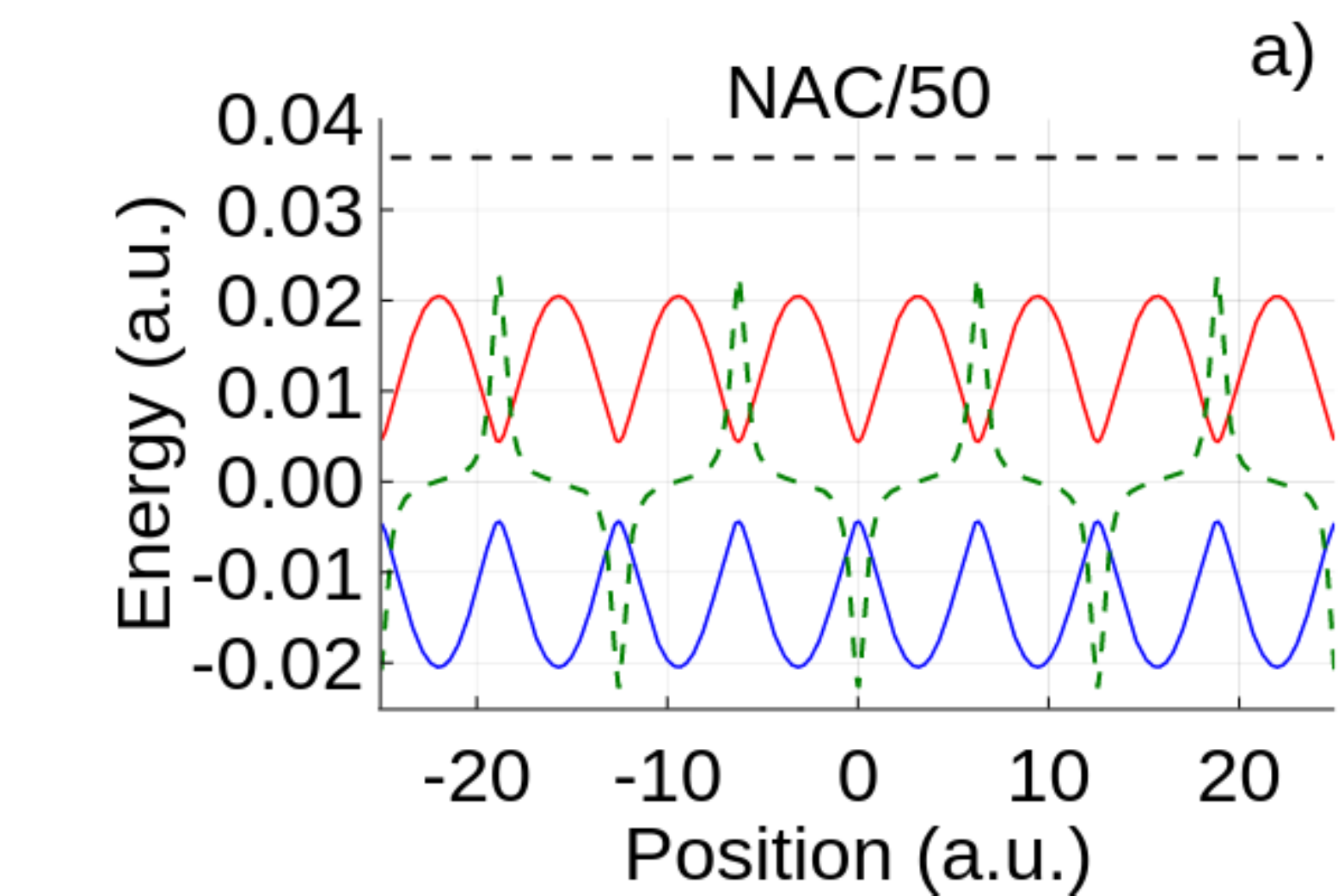}
    \includegraphics[width=7cm]{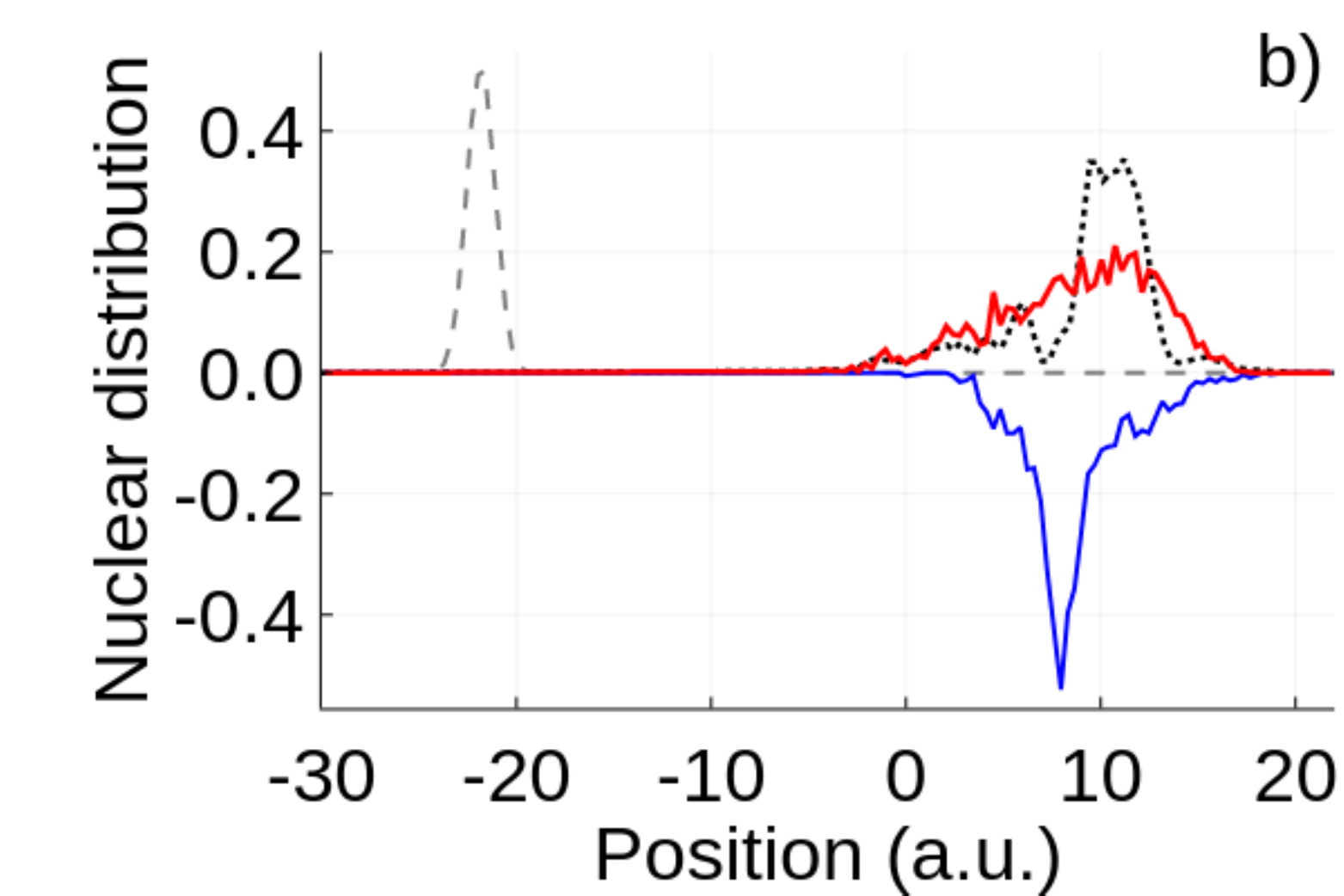}
  \caption{a) Sinusoidal model with coupling $c=0.0044$ and $\xi=0.96$:  \glspl{PES} (solid) and initial energy (horizontal dashed). b) Initial nuclear position distribution (dashed gray, its magnitude is divided by $2$, the average momentum $p_0=15$), and final nuclear position distributions:  \gls{EH} (blue, mirrored for clarity), \gls{SH} (red) and SO (black dots).}
  \label{fig:crit_bal}
\end{figure}

\paragraph{High energy regime:} 
Even setting $\xi=1$ does not break down the EH method for 
the high initial energy case (Fig.~\ref{fig:crit_bal}), all three methods produce similar dynamics. 
The rationale for this behavior is similarity in slopes of ground and excited PESs over a distance spanning several minima.  However, one can notice that the exact distribution has two parts (faster and slower) while the 
\gls{EH} distribution is centered right at their separation point. This result hints that at longer 
times the separation between two parts of the exact distribution may grow while the \gls{EH} distribution 
will be approximating their average. 

\subsection{Metallic surface model}

\begin{figure}[]
  \centering
    \includegraphics[width=7cm]{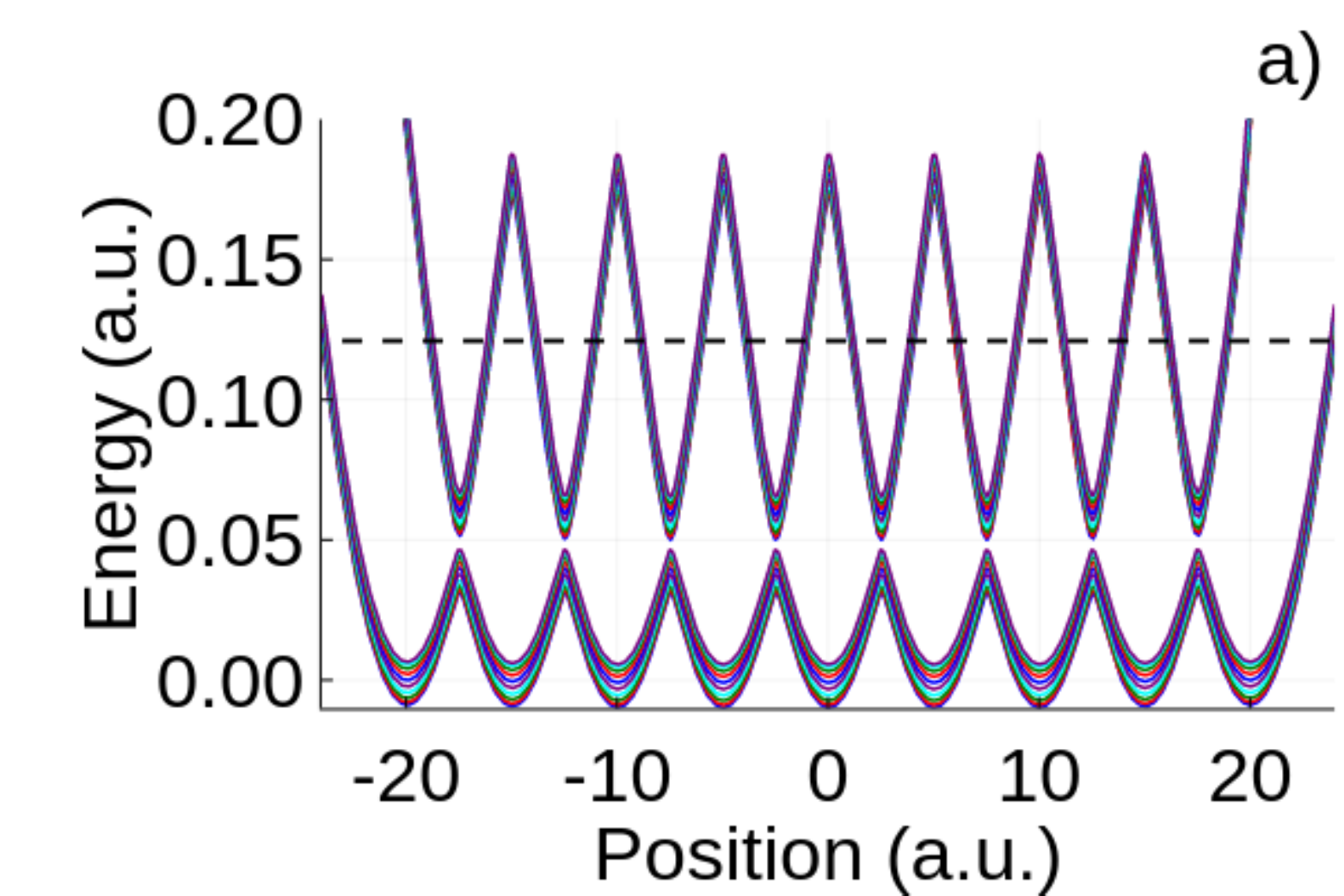}
    \includegraphics[width=7cm]{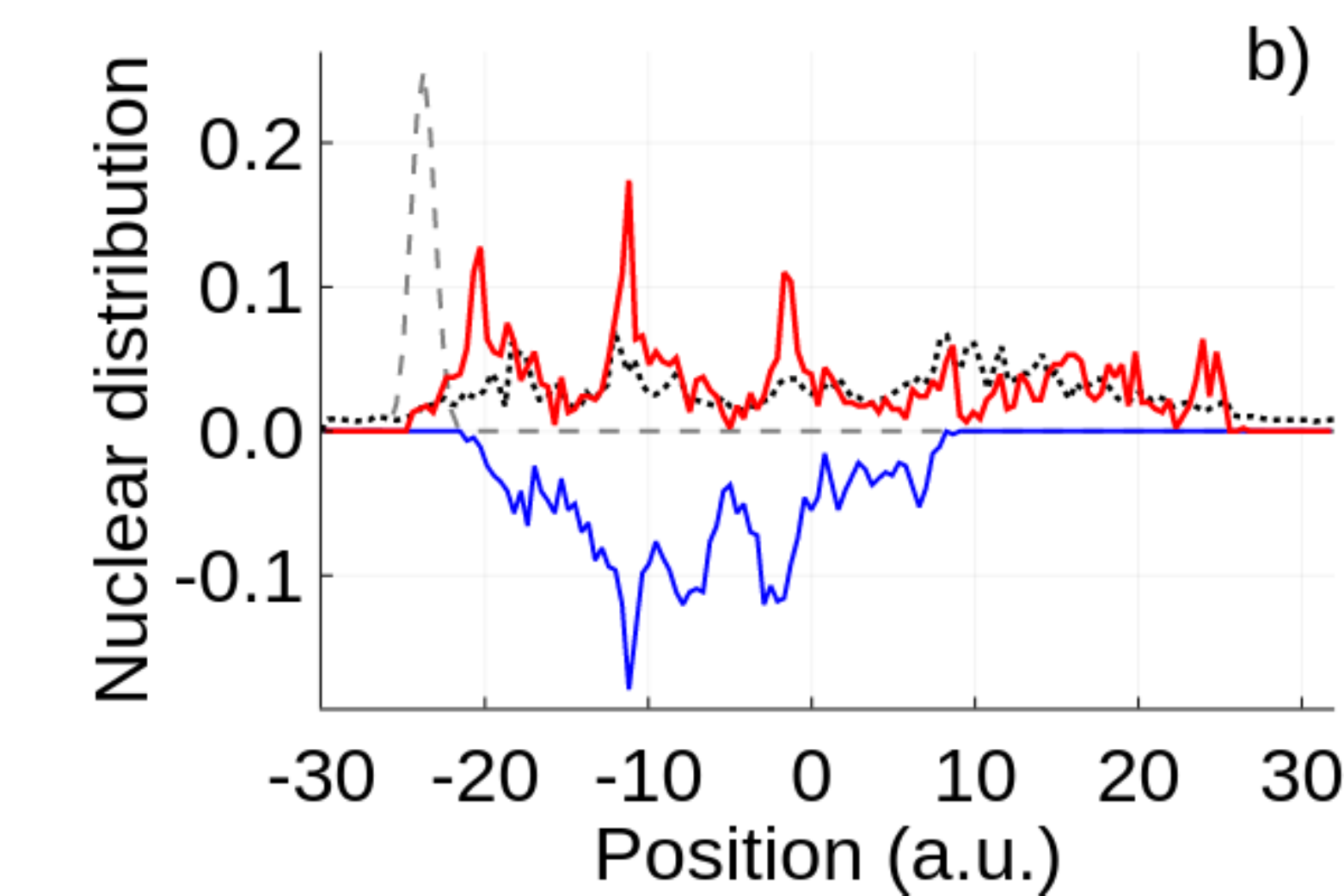}
  \caption{a) Metallic surface model with inter-layer coupling $\beta_1=0.011$ and $\xi=0.90$ (between the ground state and the first excited state of the next layer): \glspl{PES} (solid), initial energy (horizontal dashed). b) Initial (dashed gray, its magnitude is divided by $4$, zero average momentum) and final nuclear position distributions (blue for \gls{EH}, red for \gls{SH}, and black dots for \gls{SO}). \gls{EH} results were mirrored for clarity.}
  \label{fig:multi}
\end{figure}

To model dynamics on a metallic surface where more electronic states are accessible, we use the model that was originally introduced in Ref.~\citenum{ilya} for representing a chemisorbed atom on a 1D metallic chain of atoms. 
The total diabatic potential is built in three steps. First, an elementary building block is chosen as a harmonic potential 
describing the chemical bonding in an adatom-metal dimer:
\begin{equation}
V_0(R)=\frac{M\Omega^2R^2}{2},
\end{equation}
where $\Omega$ is a harmonic frequency. Second, this elementary potential is replicated by 
defining $V_{kD}(R)=V_0(R-kD)$ for $-n\leq k \leq n$, and all replicas are placed in interacting diabatic 
potential matrix
\begin{equation} \label{eq:single_layer}
V(R)=\left[ {\begin{array}{cccc}
V_{-nD}(R) & \beta_1 & \beta_2 & \dots \\
\beta_1 & V_{(-n+1)D}(R) & \beta_1 & \vdots \\
\vdots & \beta_1 & \ddots &  \\
\dots & \dots &  & V_{nD}(R)
\end{array} } \right]
\end{equation}
with $\beta_i$ coupling constants, and $D$ the distance between the atoms of the chain. 
Third, this set of diabatic potentials is replicated $m+1$ times 
vertically in the energy direction with addition of a small spatial 
shift $d$ and diabatic coupling $\alpha$ between nearest neighboring replicas. 
The scalar potential $V_{kD}$ in Eq.\eqref{eq:single_layer} is then replaced by a $(m+1)\times(m+1)$ tridiagonal matrix with diagonal elements $[\mathbb{V}_{kDd\alpha}]_{i,i}(R)=V_{kD}(R+(i-1)d)$, 
and each coupling constant becomes a $(m+1)\times(m+1)$ diagonal matrix
\begin{equation}
\mathbf{V}(R)=\left[ {\begin{array}{ccc}
\mathbb{V}_{-nDd\alpha}(R) & \bbbeta_1 & \dots \\
\bbbeta_1 & \ddots & \vdots \\
\end{array} } \right].
\end{equation}
All parameters used for this model are given in Table 1.
\begin{table}[h!]
\centering
\caption{Parameters of the metallic surface model}
\begin{tabular}{l c}
\hline
Parameter & Value (a.u.) \\ \hline
Number of atoms in a chain $2n+1$ & 9 \\
Harmonic frequency $\Omega$ &  $0.0028$  \\
Atom chain distance $D$ & $5.0$  \\
Inter-layer couplings: & \\
\hspace{2.5cm}$\beta_1$ & $0.03, 0.011$ \\
\hspace{2.5cm}$\beta_2$ & $0.02$ \\
\hspace{2.5cm}$\beta_{i>2}$ & $0.01$ \\ 
Intra-layer coupling $\alpha$ & 0.004 \\
Intra-layer offset $d$ & $1.0\times 10^{-4}$ \\
Number of layers $m+1$ & 10 \\ \hline
\end{tabular}
\label{tab:parameters}
\end{table}
To summarize, the metallic model has two types of couplings: $\alpha$, between the almost parallel PESs (intra-layer), which are to model electron-hole excitations. $\beta_i$ between the \glspl{PES} on different equilibrium positions on the surface (inter-layer), and yield adiabatic \glspl{PES} with very different slopes, as it can be seen in Fig.~\ref{fig:multi}a.

\begin{figure}[]
  \centering
    \includegraphics[width=7cm]{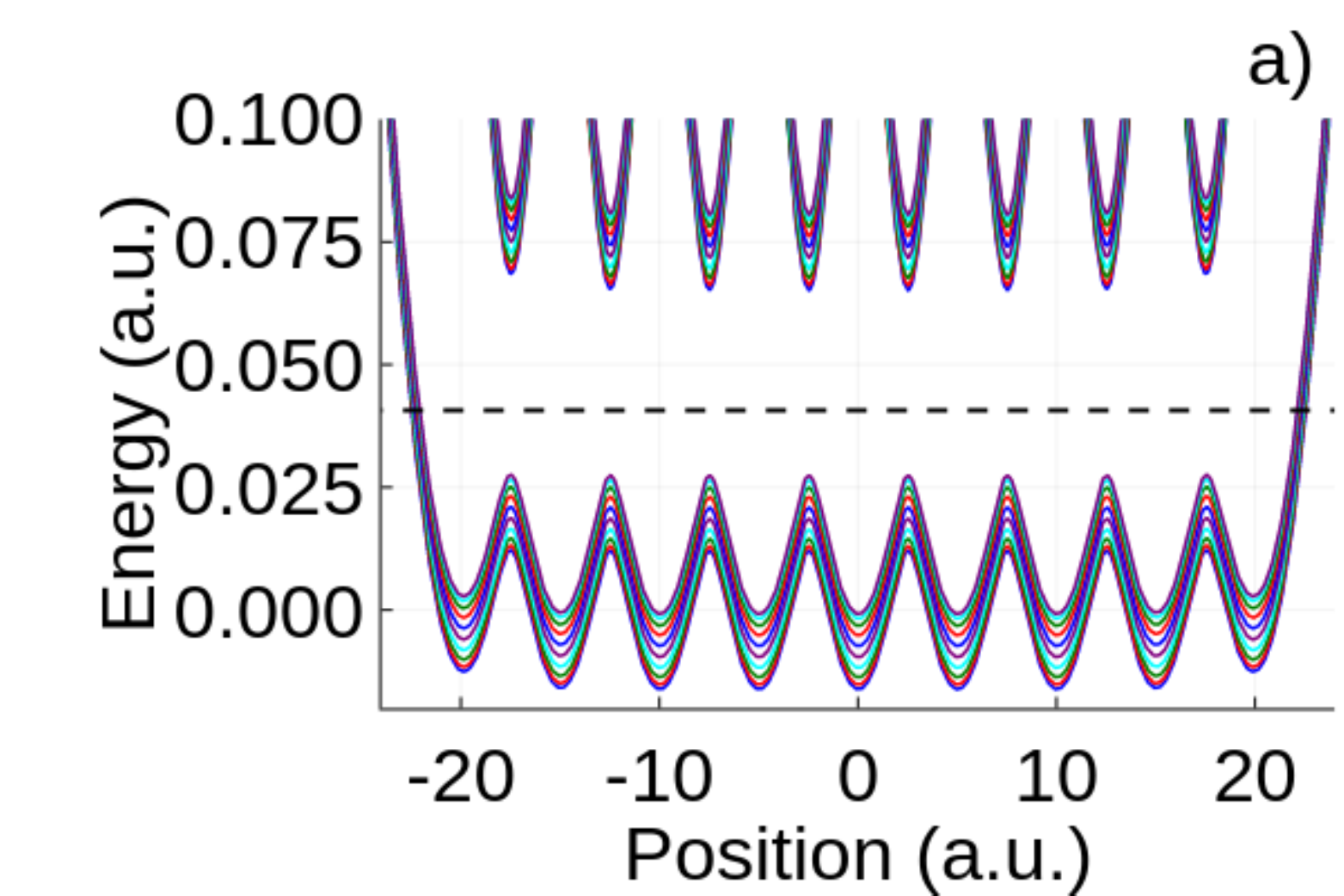}
    \includegraphics[width=7cm]{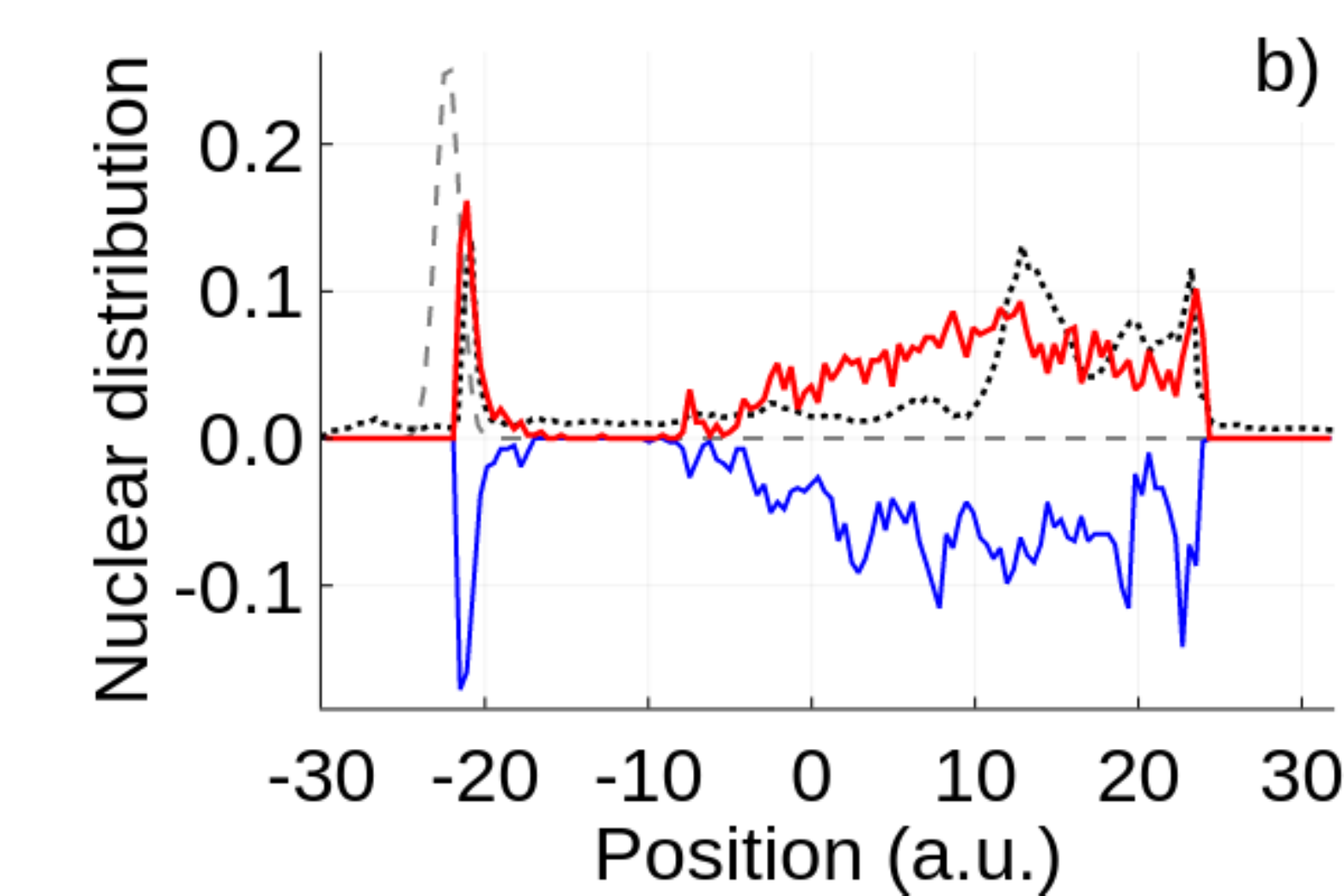}
    \includegraphics[width=7cm]{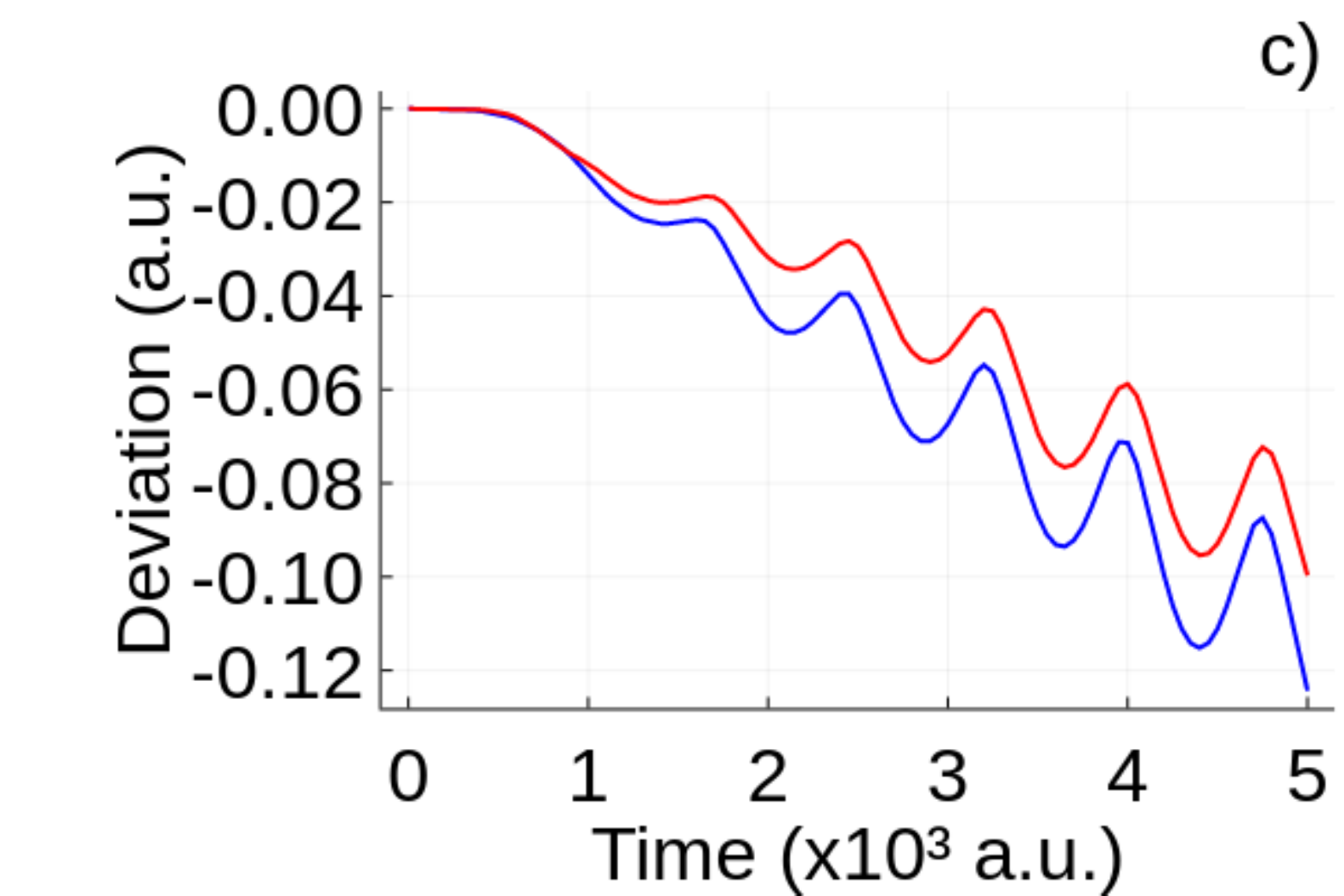}
  \caption{a) Metallic surface model, $\xi=874$ (between the ground state and the first excited state of the next layer): \glspl{PES} (solid) and initial energy (horizontal dashed). b) Initial (dashed gray, divided by $4$, zero average momentum) and final nuclear positions for nuclear trajectories (blue for \gls{EH}, red for \gls{SH}, and black dots for \gls{SO}). \gls{EH} results were mirrored for clarity. c) Deviations from a Born-Oppenheimer trajectory with
   zero initial momentum and -22 a.u. initial position.}
  \label{fig:layered_low_energy}
\end{figure}

There are two main energetic regimes which will be of interest. (1) Frictional regime: The initial energy 
is sufficient to stimulate nonadiabatic transitions between almost parallel PESs constituting the first layer 
but is insufficient for significant population of an excited adatom state corresponding to the second layer (Fig.~\ref{fig:layered_low_energy}). (2) Two-layer regime: The initial energy allows the system to have 
similar probabilities for the first and second layers (Fig.~\ref{fig:multi}).
Even higher initial energies are not expected to add new dynamical regimes because excitations to higher layers 
will only increase possible deviations of dynamics on different layers. 

{\it a. Frictional regime:} Figure~\ref{fig:layered_low_energy} shows the results for an inter-layer coupling of $\beta_1=0.03$. Both \gls{EH} and \gls{SH} yield very similar results: the electronic population has spread over the layered states, and both methods capture most of the nuclear branching. Both methods show a frictional behaviour: Fig.~\ref{fig:layered_low_energy}c shows the deviation against a trajectory on the adiabatic ground state (i.e. Born-Oppenheimer dynamics). The adiabatic Massey parameter is $\xi=874$, meaning there are no competing pathways in different layers, which is consistent with the \gls{EH} results being almost identical to the \gls{SH} ones.

{\it b. Two-layer regime:} When we used $\beta_1=0.03$, we obtained a ballistic-like motion, both \gls{EH} and \gls{SH} yield dynamics that are extremely similar to the \gls{SO} ones: we are in an adiabatic regime ($\xi=6.0$), and \gls{EH} is able to model this correctly, including the intra-layer effects. Figure \ref{fig:multi} shows the results for the $\beta_1=0.011$ regime.
As expected, \gls{EH} fails to account for the branching of the nuclear trajectories, obtaining results that are very different from the exact quantum ones, while \gls{SH} recovers results that are a lot more accurate. The inter-layer coupling is then responsible for the breakdown of \gls{EH} dynamics, and our indicator can be applied for such transitions.

\section{Conclusions}

We have analyzed break-down of the EH approach in several one-dimensional models containing multiple electronic 
states and amenable to numerically exact treatment. 
The main condition for EH failure is accessibility of several electronic PESs with different nuclear forces. 
The numerical indicator for identifying such cases, the Massey parameter, has been suggested and assessed. 
When the Massey parameter is much larger or much smaller than 1, \gls{EH} and \gls{SH} 
yield very accurate results. On the other hand, when the Massey parameter approaches 1 for several 
competing pathways, the \gls{EH} breaks down failing to simulate nuclear dynamics following one of the competing 
pathways. \gls{SH} can properly treat such cases and outperforms \gls{EH} in modeling dynamics for 
the considered models. 
The Massey parameter can be calculated using diabatic [Eq.~(\ref{eq:massey_d})] or adiabatic parameters [Eq.~(\ref{eq:massey_a})], and although the two versions are numerically somewhat different, 
they qualitatively agree in most of the cases. 
Even though, the considered models were inspired by possible PESs of periodic systems,  
our results applicable not only to dynamics on surfaces: the given criterion could be applied to 
any system with nonadiabatic transitions and localized \glspl{NAC}. 

Since friction theories are based on the \gls{EH} approach, 
the breakdown of \gls{EH} will lead to the breakdown of any friction-based method. 
When the Massey parameter approaches 1, any method that does not incorporate 
several possible paths will be unable to properly model the dynamics. 
This result is consistent with that of~\citet{shvsfric}, where it has been concluded that a 
friction approach seems to fail in the case where there are non-equivalent pathways. 
Such behavior can be explained by the breakdown of the underlining EH theory 
as opposed to an intrinsic failure of the friction approach. 

\section*{Acknowledgements}
I.L.G. is grateful to Ilya G. Ryabinkin and Rami Gherib for stimulating discussions. 
A.F.I. acknowledges the financial support from the Ontario Ministry of Research and 
Innovation through an Early Researcher Award.

\section*{Appendix A: The Landau-Zener model} \label{sec:appendix}

Here we reformulate the Massey parameter for the LZ model in the adiabatic quantities.
First, let us derive the time-derivative coupling between the adiabatic states of the LZ model.
The \gls{LZ} model potential is a time-dependent matrix in the diabatic representation
\begin{equation}
V(t)=\left[ {\begin{array}{cc}
\delta t & \Delta \\
\Delta & -\delta t \\
\end{array} } \right].
\end{equation}
The eigenfunctions (adiabatic states) 
of this potential are
\begin{equation}
\ket{\phi_1}=\begin{bmatrix}
\sin(\frac{\theta}{2}) \\
-\cos(\frac{\theta}{2})
\end{bmatrix};\ \ket{\phi_2}=\begin{bmatrix}
\cos(\frac{\theta}{2}) \\
\sin(\frac{\theta}{2})
\end{bmatrix} ,
\end{equation}
where $\tan(\theta)=\Delta/(\delta t)$. The corresponding eigenvalues are $E_1=-\sqrt{\delta^2t^2+\Delta^2}$ and $E_2=\sqrt{\delta^2t^2+\Delta^2}$, and 
the time-derivative coupling is 
\begin{align}
\tau&=\braket{\phi_1|\partial_t \phi_2}=\frac{\braket{\phi_1|\partial_t V|\phi_2}}{\omega}    \\
&=\frac{1}{\omega}
\begin{bmatrix}
\sin(\frac{\theta}{2}) \ -\cos(\frac{\theta}{2})
\end{bmatrix}
\begin{bmatrix}
\delta & 0 \\
0 & -\delta
\end{bmatrix}
\begin{bmatrix}
\cos(\frac{\theta}{2}) \\
\sin(\frac{\theta}{2})
\end{bmatrix} \\
&=\frac{\delta \sin(\theta)}{\omega},
\end{align}
where $\omega=E_2-E_1$. Accounting for $\sin(\theta)=\Delta/\sqrt{\delta^2t^2+\Delta^2}$ 
and $\omega=2\sqrt{\delta^2t^2+\Delta^2}$ we obtain
\begin{equation}
\tau=\frac{\delta\Delta}{2(\delta^2t^2+\Delta^2)}.
\end{equation}
Second, at zero time, when the particle reaches the diabatic intersection point, $\tau$ can be turned into a
quantity that will be useful for our purpose by division on $\omega$
\begin{equation} \label{eq:ad_tau}
\frac{\tau(t=0)}{\omega}=\frac{\delta\Delta}{4(\delta^2t^2+\Delta^2)^{\frac{3}{2}}}\Big |_{t=0}=\frac{\delta}{4\Delta^2}.
\end{equation}
Third, let us reformulate the diabatic Massey parameter by calculating the diabatic force difference. We can write the force matrix $F=-\nabla V$, and using the chain rule for the derivative we get
\begin{equation} \label{eq:force}
F=\frac{1}{\dot R}\dpar{V}{t}=\frac{1}{\dot R}\begin{bmatrix}
\delta & 0 \\
0 & -\delta
\end{bmatrix}.
\end{equation}
Therefore the diabatic force difference is given by $\abs{F_b-F_a}=2\delta\dot R^{-1}$. Note that the \gls{LZ} model assumes a quick transition through a crossing, meaning the nuclear velocity will not suffer any significant changes: it can be thought of as a constant, a good first order approximation if the force is acting for a short time. Replacing \eq{eq:force} into the diabatic Massey parameter:
\begin{equation}
\xi=\frac{2\pi\Delta^2}{\dot R\abs{F_b-F_a}}=\frac{\pi\Delta^2}{\delta}.
\end{equation}
Using Eq.\eqref{eq:ad_tau} we can rewrite this as
\begin{equation}
\xi=\frac{\pi}{4}\frac{4\Delta^2}{\delta}=\frac{\pi\omega}{4\tau(t=0)},
\end{equation}
obtaining the adiabatic Massey parameter shown in Eq.\eqref{eq:massey_a}.
%

\end{document}